\documentclass[12pt,letterpaper]{article}
\pdfoutput=1

\usepackage[includeheadfoot,
            marginratio={1:1,2:3}, 
            width=412pt, 
            height=688pt,]{geometry}

\usepackage{amsmath}
\usepackage{amsfonts}
\usepackage{amssymb}
\usepackage{graphicx}
\usepackage{cite}
\usepackage{ulem}
\usepackage{longtable}
\usepackage{afterpage}
\usepackage{amsthm}
\usepackage{relsize}
\usepackage[hidelinks]{hyperref}

\newcommand{\eq}[1]{\begin{equation}
                     \begin{aligned} #1 \end{aligned}
                     \end{equation}}

\newcommand{\Linf}{L$_\infty\,$}
\newcommand{\eps}{\epsilon}
\newcommand{\del}{\partial}

\DeclareMathOperator{\Unsh}{Unsh}

\allowdisplaybreaks[2]
\numberwithin{equation}{section}

\begin{document}

\normalem
\vspace*{-1.5cm}
\begin{flushright}
  {\small
  MPP-2018-223 \\
  }
\end{flushright}

\vspace{1.5cm}

\begin{center}
  {\LARGE
   On the Existence of an \Linf structure  \\[2mm]
   for the Super-Virasoro Algebra
}
\vspace{0.4cm}

\end{center}

\vspace{0.35cm}
\begin{center}
  Ralph Blumenhagen, Max Brinkmann
\end{center}

\vspace{0.1cm}
\begin{center} 
\emph{ Max-Planck-Institut f\"ur Physik (Werner-Heisenberg-Institut), \\ 
   F\"ohringer Ring 6,  80805 M\"unchen, Germany } \

\vspace{0.0cm}

 \vspace{0.3cm} 
\end{center}

\vspace{1cm}

\begin{abstract}
\noindent
The appearance of \Linf structures for  supersymmetric symmetry 
algebras in two-dimensional  conformal field theories is
investigated. Looking at the simplest  concrete example 
of the ${\cal N}=1$ super-Virasoro algebra
in detail, we investigate whether 
an extension to a  super-\Linf algebra is sufficient to capture all appearing signs.
\end{abstract}

\clearpage


\section{Introduction}
\Linf algebras were introduced to string theory in 1992 when the algebraic
structure of bosonic closed string field theory as proposed by
Zwiebach~\cite{Zwiebach92:CSFTandBV} was found to encode an \Linf
algebra. These algebras are also called strongly homotopy (sh) Lie
algebras in the mathematical literature and are generalizations of Lie
algebras. The 2-bracket of an \Linf algebra may violate the Jacobi
identity, however this failure is captured by a homotopical term  which defines a 3-bracket. The word "strongly" refers to the fact that this pattern continues, i.e. a generalized Jacobi identity for the $n$-bracket holds up to homotopical terms defining an $(n\!+\!1)$-bracket.

The connection between gauge theories and \Linf algebras was made by~\cite{Alexandrov:1995kv} using a geometric formulation of the general BV-formalism.
More recently a very tractable procedure for identifying the \Linf algebra of a gauge theory was formulated~\cite{HohmZwiebach17:LinfFieldTheory}.
There it was explicitly shown how the \Linf algebra incorporates the gauge variations, the gauge algebra
and the dynamics of the theory. In this way, not only the \Linf
structure of Yang-Mills and Chern-Simons theory but also the \Linf
structure of Double Field Theory and  Einstein gravity  was
derived. In the context of 2D  conformal field theories (CFTs), extended non-linear
classical conformal algebras, so called  classical $\mathcal{W}$ algebras,  were also
found to exhibit an \Linf structure~\cite{Blumenhagen17:WareLinf}. Quantum $\mathcal{W}$ algebras were investigated as well~\cite{Blumenhagen:2017ulg}, and a quantum \Linf algebra was proposed.
Note that $\mathcal{W}$ algebras are not gauge symmetries but infinitely many global
symmetries that can be considered to be holographically dual to
higher spin gauge symmetries.

Following these results, the logic was turned around by the \Linf bootstrap approach~\cite{Blumenhagen:2018kwq}. It was proposed that non-commutative gauge theories can be constructed iteratively by fixing the free theory and the gauge group, then imposing the \Linf relations order by order while requiring the commutative limit to flow to an ordinary gauge theory. This means that the \Linf algebra is taken as the guiding principle for obtaining new theories. With this method, derivative and curvature corrections to the equations of motion can be bootstrapped in an algebraic way. 
As it requires quite simple mathematics in the actual computation (although the equations get exponentially complicated in higher orders), this is an extremely powerful way to obtain general deformations of gauge theories.
The question of uniqueness of the bootstrap was subsequently addressed and led to a theorem of equivalence between the two well established concepts of \Linf quasi-isomorphisms and Seiberg-Witten maps~\cite{Blumenhagen18:LinfUnique}.

All of this was developed using purely bosonic theories. A natural question in this context is if there exists an extension of the formalism to supersymmetric theories. 
While \Linf algebras and superalgebras have appeared in the same title before~\cite{Cederwall:2018aab}, the theory considered there does not contain physical fermions.
It might be expected that we need to extend \Linf algebras to
super-\Linf algebras. 
In this note  we investigate this question in the context of supersymmetric
extensions of 2D conformal symmetries. To make the first steps we consider a
very simple prototype example, namely the ${\cal N}=1$ super-Virasoro
algebra.

This letter is structured as follows:
First we will briefly introduce the necessary concepts and notations
of (super-)\Linf and super-Virasoro algebras. We then naively follow
the calculations presented in~\cite{Blumenhagen17:WareLinf} to derive
\Linf-like maps from the symmetry variations of the chiral fields. 
Once the maps are defined, we can discuss the graded vector space
that the algebra should be defined on. We will see that with a 
super-extension of the \Linf algebra it is possible to capture 
the extra $\mathbb Z_2$ grading present in super-${\cal W}$ algebras. 
As the main result, we find that the \Linf formalism for bosonic theories
 proposed in \cite{HohmZwiebach17:LinfFieldTheory}
is also valid for supersymmetric symmetry algebras by simply extending the \Linf to a super-\Linf algebra.

\section{Preliminaries}
In this section we present a brief  introduction to \Linf algebras and
the super-Virasoro algebra. We will keep it short and only focus on
the material that is compulsory  for  the following.
For further information on \Linf algebras we refer to
\cite{HohmZwiebach17:LinfFieldTheory,Lada93:SHLieAlgPhys,Lada94:shLieAlg}
and for more on CFTs and the super-Virasoro algebra see
e.g. \cite{Simmons-Duffin16:ConfBootstrap,GKO86:sVirRepr,Figueroa90:sW-alg}
or \cite{Blumenhagen09:IntroCFT}  for a textbook introduction.

\subsection{Basics of \Linf algebras}\label{basicsLinf}
There are multiple equivalent definitions of \Linf algebras \cite{HohmZwiebach17:LinfFieldTheory,Lada93:SHLieAlgPhys,Lada94:shLieAlg}.
We will use the so-called $\ell$-picture. While this picture has the
disadvantage of many more sign factors compared to the $b$-picture,
they  allow us to give the first maps a nice interpretation.  

Let $X$ be a $\mathbb{Z}$-graded vector space 
\eq{
X=\underset{n\in\mathbb{Z}}{\bigoplus}X_n \,,
} 
where elements of $X_n$ have degree $n$. We will use $x_1,x_2,...$ to
represent arbitrary vectors of fixed degree in $X$, $x_i \in X_n$. 
Here $i$ is just an arbitrary label to distinguish variables while $n$ is the degree
$\deg(x_i)=n$. 
The degree will appear in sign factors, and we will omit the 'deg' label $(-1)^{\deg(x_1)\deg(x_2)}\equiv(-1)^{x_1x_2}$ where confusion is impossible.

Let $\{\ell_n:X^{\otimes n}\rightarrow X\}_{n>0}$ be graded anti-commutative multilinear maps of degree $\deg(\ell_n)=n-2$ acting on elements of homogeneous degree $x_i\in X$: 
\eq{
\ell_n(x_1,...,x_i,x_j,...,x_n)&=(-1)^{1+x_i x_j} \ell_n(x_1,...,x_j,x_i,...,x_n)\;, \\ 
\deg(\ell_n(x_1,...,x_n))&=\sum_{i=1}^{n}\deg(x_i)+n-2 \;.
}
Then $\left(X,\{\ell_n\}\right)$ is an \Linf algebra if the maps satisfy the following defining
relations for all $n>0$:
\eq{\label{lDefRel}
\mathcal{J}_{n} :=\sum_{\substack{l,k\geq0\\l+k=n}} \sum_{\sigma\in \Unsh(k,l)}& (-1)^{kl}\;(-1)^{\sigma}\;  \eps(\sigma;x)\; 
\\
&\times\ell_{l+1}(\ell_k(x_{\sigma(1)},...,x_{\sigma(k)}),x_{\sigma(k+1)},...,x_{\sigma(n)})\;=\;0
}
where $(-1)^\sigma$ is the sign of the permutation and gives a positive (negative) sign for even (uneven) permutation $\sigma$. 
The sum runs over all unshuffles $\sigma\in\Unsh(k,l)$. Unshuffles correspond to inequivalent partitions of a set with $n$ elements into two subsets of length $k$ and $l$. Unshuffles are equivalent if the subsets contain the same elements respectively, independent of their order. For example the partition $(1|23)$ is equivalent to $(1|32)$ and inequivalent to $(2|13)$.
The Koszul sign $\eps(\sigma;x)$ is defined as the sign needed to rearrange a graded commutative algebra $\Lambda(x_1,...,x_n)$:
\eq{\label{KoszulSign}
x_{\sigma(1)}\wedge...\wedge x_{\sigma(n)} = \eps(\sigma;x) x_1\wedge...\wedge x_n \,.
}
The \Linf defining relations can be written schematically as 
\eq{
0=\ell_1\ell_1 \;,\quad 0= \ell_1\ell_2-\ell_2\ell_1 \;,\quad 0=\ell_1\ell_3+\ell_2\ell_2+\ell_3\ell_1 \;,\quad... \,.
}
Explicitly, the \Linf relations for $n=1,2,3$ read:
\eq{\label{lDefRelExpl}
0&=&\mathcal{J}_1&=\ell_1(\ell_1(x)) \,,
\\[5pt]
0&=&\mathcal{J}_2 &= \ell_1(\ell_2(x_1,x_2)) - \ell_2(\ell_1(x_1),x_2)) - (-1)^{x_1} \ell_2(x_1,\ell_1(x_2)) \,,
\\[5pt]
0&=&\mathcal{J}_3 &= \ell_1(\ell_3(x_1,x_2,x_3)) + \ell_3(\ell_1(x_1),x_2,x_3)  \\
&&&\phantom{=} + (-1)^{x_1} \ell_3(x_1,\ell_1(x_2),x_3) + (-1)^{x_1+x_2} \ell_3(x_1,x_2,\ell_1(x_3))\\
&&&\phantom{=} + \ell_2(\ell_2(x_1,x_2),x_3) + (-1)^{x_1(x_2+x_3)} \ell_2(\ell_2(x_2,x_3),x_1) \\
&&&\phantom{=} + (-1)^{x_3(x_1+x_2)} \ell_2(\ell_2(x_3,x_1),x_2) \,.
}

The first two equations show that $\ell_1$ is nilpotent and acts as a derivation of $\ell_2$. The third equation contains two more important features. The first two lines show that $\ell_1$ fails to be a derivation of $\ell_3$. However the last two lines are the (graded) Jacobi identity for $\ell_2$ whose failure to hold is controlled by the first two lines. Mathematically, $\ell_3$ is a chain homotopy. Thus $\ell_2$ satisfies the Jacobi identity up to homotopy exact terms. This pattern continues for higher $\ell$-brackets: for $n$ inputs a generalized Jacobi identity of the schematic form $\sum_{i,j>1}\pm\ell_i\ell_j$ will hold up to $\ell_1\ell_n\pm\ell_n\ell_1$ terms \cite{Lada93:SHLieAlgPhys}.

A super-\Linf algebra is an \Linf algebra defined on a super vector
space~\cite{nlab:SLinf}. This means one adds an underlying
$\mathbb{Z}_2$ grading to the algebra by allowing the vector spaces of homogeneous degree to be a direct sum of a Grassmann even (bosonic) and a Grassmann odd (fermionic) vector space:
\eq{\label{sLinf1}
X=\underset{n\in\mathbb{Z}}{\bigoplus}X_n \,, \qquad
X_n = X_n^{\text{bos}} \oplus X_n^{\text{ferm}} \,.
} 
Equivalently, the super-\Linf algebra is defined on a $\mathbb{Z}\times\mathbb{Z}_2$ graded (Grassmann even) vector space $X$ such that an element of homogeneous degree $x_i\in X$ is equipped with two labels $(n_i,s_i)$ with $n_i\in \mathbb{Z}$ and $s_i\in\{0,1\}$,
\eq{\label{sLinf2}
X=\underset{\substack{n\in\mathbb{Z} \\s\in\{0,1\}}}{\bigoplus}X_{n,s} \,.
} 
The multilinear maps  are then anti-commutative with respect to both $n$ and $s$,
\eq{\label{supergradingsymmetry}
\ell_n(x_1,...,x_i,x_j,...,x_n)&=(-1)^{1+n_i n_j +s_is_j} \ell_n(x_1,...,x_j,x_i,...,x_n) \,.
}
The degree of $x_i$ is therefore to be given by two integers $(n_i,s_i)$ instead of only one number. 
The \Linf defining relations are also modified by taking the Koszul sign with
respect to both $n$- and $s$-grading. This ensures that both descriptions are equivalent.
In the following, we will refer to the fist description as an \Linf
algebra over a super vector space and to the second description as a
super-\Linf algebra, keeping in mind that both are equivalent.

\subsection{\Linf algebras and extended conformal symmetries}
We will follow the dictionary for \Linf algebras in gauged theories developed
in~\cite{HohmZwiebach17:LinfFieldTheory}. When applying this to 
bosonic extensions of the conformal symmetry,
the vector space consists of only two terms
\eq{
X =  \underset{\substack{\rotatebox[origin=c]{90}{$\in$}\\ \mathlarger{\epsilon}}}{X_0}
\oplus 
\underset{\substack{\rotatebox[origin=c]{90}{$\in$}\\
    \mathlarger{W}}}{X_{-1}}  
}
with chiral fields $W(z)$ and symmetry  parameters $\epsilon(z)$. The fields transform under infinitesimal  transformations as
\eq{\label{GaugeTrafo}
\delta_\epsilon W &= \sum_{n\geq 0}\frac{1}{n!}(-1)^{\frac{n(n-1)}{2}}\ell_{n+1}(\epsilon,W^n)\\
&=\ell_1(\epsilon)+\ell_2(\epsilon,W)-\frac{1}{2}\ell_3(\epsilon,W,W)-\frac{1}{3!}\ell_4(\epsilon,W,W,W)+...
}
and the symmetry algebra is given in terms of the \Linf algebra as
\eq{\label{GaugeCommutator}
\left[\delta_{\epsilon_1},\delta_{\epsilon_2}\right]W &= \delta_{-\mathcal{C}(\epsilon_1,\epsilon_2,W)}W \,,
\\[5pt]
\mathcal{C}(\epsilon_1,\epsilon_2,W) &= \sum_{n\geq0}\frac{1}{n!}\ell_{n+2}(\epsilon_1,\epsilon_2,W^n)
\\
&= \ell_2(\epsilon_1,\epsilon_2) + \ell_3(\epsilon_1,\epsilon_2,W)-\frac{1}{2}\ell_4(\epsilon_1,\epsilon_2,W^2)-... \,.
}

\subsection{The super-Virasoro algebra}
The super-Virasoro algebra we consider is the $\mathcal{N}=1$ Neveu-Schwarz supersymmetric extension of the Virasoro algebra. The Virasoro algebra contains the generators of infinitesimal conformal transformations of $2D$ conformal field theories and is thus related to the energy-momentum tensor. 

The super-Virasoro algebra
consists of the energy-momentum tensor $L(z)$ with conformal dimension $(h=2)$ and a fermionic primary field $G(z)$ of dimension $(h=\frac{3}{2})$ as its superpartner. The mode expansions of the fields are given by
\eq{
L(z) = \sum_{m\in \mathbb{Z}}L_m \, z^{-m-2}  \;,\quad G(z)=\sum_{r\in\mathbb{Z}+\frac{1}{2}} G_r \, z^{-r-\frac{3}{2}}
}
with the modes of $G$ being half-integers. Note that $G$ as well as $G_r$ are fermionic and thus anticommuting. The super-Virasoro algebra is then given as 
\eq{\label{sVir}
[L_m , L_n ] &= (m-n)L_{m+n} + \frac{c}{12}(m^3-m)\delta_{m+n,0} \,, \\
[L_m , G_r ] &= \left(\frac{m}{2}-r\right) G_{m+r} \,, \\
\{G_r,G_s \} &= 2 L_{r+s} + \frac{c}{3}\left(r^2-\frac{1}{4}\right)\delta_{r+s,0} \,,
}
with the anticommutator appearing because $G_r$ are Grassmann odd
\cite{GKO86:sVirRepr}.  Note that in the here considered classical
case, these are actually super Poisson-brackets.

In the following, for computational convenience we will employ the OPE representation of the algebra \cite{Figueroa90:sW-alg}:
\eq{
L(z)\,L(w) &= \frac{c/2}{(z-w)^4} + \frac{2 L(w)}{(z-w)^2} + \frac{\del_w L(w)}{z-w} +... \,, \\[2pt]
L(z)\,G(w) &= \frac{\frac{3}{2} G(w)}{(z-w)^2} + \frac{\del_w G(w)}{z-w}+... \,, \\[2pt]
G(z)\,G(w) &= \frac{\frac{2}{3} c}{(z-w)^3} + \frac{2 L(w)}{z-w} +... \,.
}

\section{The super-Virasoro algebra as an \Linf algebra}
The \Linf algebra for field theories was motivated by bosonic closed
string field theory and was shown to hold  also for bosonic gauge
theories~\cite{HohmZwiebach17:LinfFieldTheory} and bosonic symmetry algebras~\cite{Blumenhagen17:WareLinf}. It is a priori
not clear  whether  the super-Virasoro algebra also features a (super-)\Linf structure, and if so whether a super-extension of the \Linf algebra is strictly necessary.
In this section we
proceed by carefully deriving \Linf  products from the symmetry
variations of the chiral fields and their closed algebra assuming that
they fit into the relations \eqref{GaugeTrafo} and
\eqref{GaugeCommutator}.
This provides sufficient information  to fix the degree of the
vector spaces.

\subsection{The \Linf products}
The variations of the fields $X \in \{L,G\}$ with respect to the symmetry generated by the modes of $Y \in \{L,G\}$ are given by
\eq{
\delta_{\eps^Y}X(w)=\frac{1}{2\pi i} \; \oint_{\mathcal{C}(0)} d z\; \eps^Y(z) [Y(z),X(w)]_\pm \,.
}
Here $[X,Y]_\pm$ denotes  the anticommutator if both $X$ and $Y$ are fermionic, and the commutator in all other cases. Using radial ordering one can rewrite this as an integral over the OPE around $w$
\eq{
\delta_{\eps^Y}X(w)=\frac{1}{2\pi i} \; \oint_{\mathcal{C}(w)} d z\; \eps^Y(z) \Big(Y(z)X(w)\Big)_{\text{OPE}}
} 
where the contour integral extracts the singular part of the OPE.
In this way we find  the four symmetry variations
\eq{
\delta_{\eps^L}L &= \frac{c}{12}\del^3 \eps^L + 2L\,\del \eps^L + \eps^L\,\del L \,, \\[3pt]
\delta_{\eps^G}G &= \frac{c}{3}\del^2\eps^G + 2 \eps^G\, L  \,,\\[3pt]
\delta_{\eps^L}G &= \frac{3}{2}G \,\del \eps^L + \del G \,\eps^L \,, \\[3pt]
\delta_{\eps^G}L &= -\frac{3}{2}\del\eps^G\, G - \frac{1}{2}\eps^G \,\del G \,.
}
For the following it is important to note that,
as the variation of a bosonic (fermionic) field should also be bosonic
(fermionic), $\eps^L$ must be Grassmann even and $\eps^G$ Grassmann
odd. One must therefore be careful with extra  signs when computing
the symmetry  algebra and  the \Linf relations.

Using  \eqref{GaugeTrafo}, one can directly read off the
nontrivial products with one symmetry  parameter
\eq{\label{sVirGaugeBrackets}
\ell_1^L(\eps^L)&=\frac{c}{12}\del^3\eps^L \,, \\[2pt]
\ell_1^G(\eps^G)&=\frac{c}{3}\del^2\eps^G \,, \\[2pt]
\ell_2^L(\eps^L,L)&=2 \del\eps^L L + \eps^L \del L \,, \\[2pt]
\ell_2^G(\eps^L,G)&=\frac{3}{2} \del\eps^L G +\eps^L\del G \,, \\[2pt]
\ell_2^G(\eps^G,L)&=2 \eps^G L \,, \\[2pt]
\ell_2^L(\eps^G,G)&=-\frac{3}{2}\del\eps^G G - \frac{1}{2}\eps^G\del G \,.
}
The superscript of the products indicates the type of field the products map to. Next we have 6 equations to calculate the commutator of two symmetry  transformations
\eq{
\big[\delta_{\eps^X},\delta_{\tilde{\eps}^Y} \big]Z &= \delta_{-\eps_{(XY)}}Z \;,\qquad X,Y,Z\in \{L,G\} \,.
}
However we expect the algebra to close, i.e. the commutator of two
transformations is again a transformation with some new
parameter. This means it should not  matter on which field we act, the
commutator should give the same new parameter. This is indeed the case
and one obtains
\eq{\label{sVirComm}
\eps_{(LL)}^L &= \eps^L \del \tilde{\eps}^L - \del \eps^L \tilde{\eps}^L \,, \\
\eps_{(GG)}^L &= -2 \eps^G \tilde{\eps}^G \,, \\
\eps_{(LG)}^G &= \eps^L\del\eps^G - \frac{1}{2}\del\eps^L\eps^G \,,
}
where the superscript denotes the kind of transformation the resulting
symmetry parameter is associated to. As expected, both the commutator of
two bosonic as well as of two fermionic transformations give a bosonic
transformation, while the commutator of a bosonic and  a fermionic
transformation gives a fermionic transformation. Using
(\ref{GaugeCommutator}) we can directly identify 
the \Linf products:
\eq{\label{sVirCommBrackets}
\ell_2^{\eps^L}(\eps^L,\tilde{\eps}^L) &= \eps^L \,\del \tilde{\eps}^L - \del \eps^L \,\tilde{\eps}^L \,,\\[5pt]
\ell_2^{\eps^L}(\eps^G,\tilde{\eps}^G) &= -2 \eps^G \,\tilde{\eps}^G \,, \\
\ell_2^{\eps^G}(\eps^L,\eps^G) &= \eps^L\,\del\eps^G - \frac{1}{2}\del\eps^L\,\eps^G \,.
}
We have now read off all the \Linf products of the super-Virasoro
algebra. In the next section we will find out what vector space actually underlies them.

\subsection{The graded vector space}

Having two distinct fields $L$, $G$ as well as two symmetry parameters $\eps^L$, $\eps^G$ in the theory, 
we can treat the space as effectively four-dimensional.
Assuming the degrees are mapped correctly, we have an equation for the relative degree of the vector spaces for each map defined in \eqref{sVirGaugeBrackets} and \eqref{sVirCommBrackets}:
\eq{
 \; \ell_n^X(A,B,...)\ne 0\quad \Rightarrow \quad |X| = |A|+|B|+...+n-2 \,.
}
With this data we can fix the graded vector space. Consider for example the following set of equations:
\eq{
\ell_1^L(\eps^L): && |L|&=|\eps^L|-1 \,, \\
\ell_1^G(\eps^G): && |G|&=|\eps^G|-1 \,, \\
\ell_2^L(\eps^L,L): && |L| &= |\eps^L|+|L| \,, \\
\ell_2^{\eps^L}(\eps^G,\tilde{\eps}^G): && |\eps^L|&= |\eps^G| \times 2 \,.
}

From the first and third equation, clearly the bosonic components have the expected degree $|L|=-1$ and $|\eps^L|=0$. The last equation fixes $|\eps^G|=0$ and with the second equation the final degree is also fixed. All in all we end up with
\eq{
 |L|=|G|=-1 \; ,\quad |\eps^L|=|\eps^G|=0 \,.
}

Thus we can write the total vector space as a graded vector space such that the homogeneous graded subspaces are again a direct sum of a bosonic and a fermionic space
\eq{\label{VectorSpace}
X=X_0\oplus X_{-1}\;,\quad X_{i\in\{0,-1\}} = X_i^{\text{bos}}\oplus X_i^{\text{ferm}} \,,
}
 with elements
\eq{
L &\in X_{-1}^\text{bos},\quad & G&\in X_{-1}^\text{ferm}, \\
\eps^L &\in X_{0}^\text{bos}, & \eps^G &\in X_{0}^\text{ferm}.
}
This is exactly of the form \eqref{sLinf1}.
We can also explicitly compute the symmetry properties of the $\ell_2$-products
\eq{
\ell_2^{\eps^L}(\eps^L,\tilde{\eps}^L)&=
\eps^L \,\del \tilde{\eps}^L - \del \eps^L \,\tilde{\eps}^L=
-\Big( \tilde{\eps}^L \,\del \eps^L -\del \tilde{\eps}^L \,\eps^L\Big)=
-\ell_2^{\eps^L}(\tilde{\eps}^L,\eps^L) \,, \\
\ell_2^{\eps^L}(\eps^G,\tilde{\eps}^G) &= -2 \eps^G \,\tilde{\eps}^G =
2 \tilde{\eps}^G\, \eps^G=-\ell_2^{\eps^L}(\tilde{\eps}^G,\eps^G) \,,
}
where in the second line we have used that $\eps^G$ is Grassmann odd.
Both $\ell_2$ products are antisymmetric which is just the
symmetry property already contained in the \Linf algebra
without introducing any further $\mathbb{Z}_2$ grading to the maps. 
This indicates that we have in fact found an \Linf algebra over a super vector space that describes the super-Virasoro algebra, 
assuming of course the \Linf relations are found to hold. 

We will proceed with checking this in the next section.

\subsection{Checking the \Linf relations}
Since we only have fields with degree $0$ and $-1$ and no additional $\mathbb{Z}_2$ grading on the maps, we expect the usual \Linf relations \eqref{lDefRel} to hold. There is no $X_{-2}$ space, so the $\mathcal{J}_1$ relation $\ell_1\ell_1=0$ is trivially satisfied. As per the argument in~\cite{Blumenhagen17:WareLinf}, we only need to consider $\mathcal{J}_2$ and $\mathcal{J}_3$ with 2 or 3 gauge parameters as inputs\footnotemark.

\footnotetext{
Although the \Linf relations are defined on the homogeneous graded spaces $X_i$ with $X_i=X_i^\text{bos}\oplus X_i^\text{ferm}$, one can choose a basis with homogeneously Grassmann even and odd vectors. Then with linearity of the \Linf products it suffices to show that the \Linf relations hold for any combination of Grassmann even/odd elements.}

The computations are lengthy but elementary. We start with $\mathcal{J}_2$, which we need to compute on combinations of two symmetry parameters. 
\begin{align*}
\mathcal{J}_2(\eps^L_1,\eps^L_2)
&
=\ell_1\Big(\eps^L_1\del\eps^L_2-\del\eps^L_1\eps^L_2\Big)-
\ell_2\left(\frac{c}{12}\del^3\eps^L_1,\eps^L_2\right)-
\ell_2\left(\eps^L_1,\frac{c}{12}\del^3\eps^L_2\right)
\\
&=\frac{c}{12}\Big(
\del^3\left(\eps^L_1\del\eps^L_2-\del\eps^L_1\eps^L_2\right)+
2\del^3 \eps^L_1 \del \eps^L_2 + \eps^L_2\del^4\eps^L_1  -
2\del^3 \eps^L_2 \del \eps^L_1 - \eps^L_1\del^4\eps^L_2
\Big)
\\
&=0.
\\ \\
\mathcal{J}_2(\eps^G_1,\eps^G_2)
&= \ell^L_1\left(\frac{1}{2}\eps^G_1 \eps^G_2\right)-\ell_2^L\left(\frac{c}{12}\del^2\eps^G_1,\eps^G_2\right) -\ell_2^L\left(\eps^G_1, \frac{c}{12}\del^2\eps^G_2\right)
\\
&=\frac{c}{24}\Big(\del^3\left(\eps^G_1\eps^G_2\right)
+\del\eps^G_2 \del^2\eps^G_1 +\eps^G_2\del^3\eps^G_1 -\del\eps^G_1\del^2\eps^G_2-\eps^G_1\del^3\eps^G_2 \Big)
\\
&=0\,.
\\ \\
\mathcal{J}_2(\eps^L,\eps^G)
&= \ell_1^G\left(\eps^L\del\eps^G-\frac{1}{2}\eps^G\del\eps^L\right) -\ell_2^G\left(\frac{c}{12}\del^3\eps^L,\eps^G\right) -\ell_2^G\left(\eps^L,\frac{c}{12}\del^2\eps^G\right) 
\\
&=\frac{c}{12}\Bigg(\del^2\left(\eps^L\del\eps^G-\frac{1}{2}\eps^G\del\eps^L\right) +\frac{1}{2} \eps^G\del^3\eps^L-\frac{3}{2}\del^2\eps^G\del\eps^L-\eps^L\del^3\eps^G\Bigg)
\\
&=0\,.
\end{align*}

With this all $\mathcal{J}_2$ relations are satisfied. In the second equation we have used that $\eps^G$ fields anticommute. 
Since there are no $\ell_3$ products, the $\mathcal{J}_3$ relations reduce to graded Jacobi identities
\eq{ 
\ell_2(\ell_2(x,y),z)+(-1)^{x(y+z)}\ell_2(\ell_2(y,z),x)+(-1)^{(x+y)z}\ell_2(\ell_2(z,x),y).
}
Next we compute the $\mathcal{J}_3$ relations with two gauge parameters. 

\begin{align*}
 \mathcal{J}_3(\eps^L_1,\eps^L_2,L)&=
 2L\del\left(\eps^L_1\del\eps^L_2-\del\eps^L_1\eps^L_2\right)+\left(\eps^L_1\del\eps^L_2-\del\eps^L_1\eps^L_2\right)\del L
\\
&\phantom{=}+2\left(2L\del\eps^L_1+\eps^L_1\del L\right)\del\eps^L_2+\eps^L_2\del\left(2L\del\eps^L_1+\eps^L_1\del L\right)
\\
&\phantom{=}-2\left(2L\del\eps^L_2+\eps^L_2\del L\right)\del\eps^L_1-\eps^L_1\del\left(2L\del\eps^L_2+\eps^L_2\del L\right)
\\
&=0.
\\ \\
\mathcal{J}_3(\eps^L_1,\eps^L_2,G)&=
\frac{3}{2}G\del\left(\eps^L_1\del\eps^L_2-\del\eps^L_1\eps^L_2\right)+\left(\eps^L_1\del\eps^L_2-\del\eps^L_1\eps^L_2\right)\del G
\\
&\phantom{=}+\frac{3}{2}\left(\frac{3}{2}G\del\eps^L_1+\eps^L_1\del G\right)\del\eps^L_2 + \eps^L_2\del\left(\frac{3}{2}G\del\eps^L_1+\eps^L_1\del G\right)
\\
&\phantom{=}-\frac{3}{2}\left(\frac{3}{2}G\del\eps^L_2+\eps^L_2\del G\right)\del\eps^L_1 - \eps^L_1\del\left(\frac{3}{2}G\del\eps^L_2+\eps^L_2\del G\right)
\\
&=0.
\\ \\
\mathcal{J}_3(\eps^G_1,\eps^G_2,L) &=
L\del\left(\eps^G_1\eps^G_2\right)+\frac{1}{2}\eps^G_1\eps^G_2\del L -
\frac{3}{4}\del\eps^G_1\eps^G_2L \\
& \phantom{=} -\frac{1}{4}\eps^G_1\del\left(\eps^G_2L\right)+
\frac{3}{4}\del\eps^G_2\eps^G_1L+\frac{1}{4}\eps^G_2\del\left(\eps^G_1L\right)
\\
&=0.
\\ \\
\mathcal{J}_3(\eps^G_1,\eps^G_2,G) &=
\frac{3}{4}G\del\left(\eps^G_1\eps^G_2\right)+\frac{1}{2}\eps^G_1\eps^G_2\del G \\
&\phantom{=}-\frac{1}{4}\eps^G_1\left(3\del\eps^G_2G+\eps^G_2\del G\right) +\frac{1}{4}\eps^G_2\left(3\del\eps^G_1G+\eps^G_1\del G\right)
\\
&=0.
\\ \\
\mathcal{J}_3(\eps^L,\eps^G,L) &=
\frac{1}{2}\left(\eps^L\del\eps^G-\frac{1}{2}\eps^G\del\eps^L\right)L 
- \frac{3}{4}\eps^GL\del\eps^L
\\
&\phantom{=}-\frac{1}{2}\eps^L\del\left(\eps^GL\right)
+ \eps^G\left(L\del\eps^L+\frac{1}{2}\eps^L\del L\right)
\\
&=0.
\\ \\
\mathcal{J}_3(\eps^L,\eps^G,G) &=
\frac{3}{2}\del\left(\eps^L\del\eps^G-\frac{1}{2}\eps^G\del\eps^L\right) G +\frac{1}{2}\left(\eps^L\del\eps^G-\frac{1}{2}\eps^G\del\eps^L\right)\del G
\\
&\phantom{=}-\left(3\del\eps^GG+\eps^G\del G\right)\del\eps^L - \eps^L\del\left(\frac{3}{2}\del\eps^GG+\frac{1}{2}\eps^G\del G\right)
\\
&\phantom{=}+\frac{3}{2}\del\eps^G\left(\frac{3}{2}G\del\eps^L+\eps^L\del G\right) +\frac{1}{2}\eps^G\del\left(\frac{3}{2}G\del\eps^L+\eps^L\del G\right)
\\
&=0.
\end{align*}

Finally there are four $\mathcal{J}_3$ relations with three symmetry parameters left to consider.

\begin{align*}
 \mathcal{J}_3(\eps^L_1,\eps^L_2,\eps^L_3)&=
 \left(\eps^L_1\del\eps^L_2-\del\eps^L_1\eps^L_2\right)\del\eps^L_3
-\del\left(\eps^L_1\del\eps^L_2-\del\eps^L_1\eps^L_2\right)\eps^L_3
+\text{cyclic} \\
&=0.
\\ \\
\mathcal{J}_3(\eps^L_1,\eps^L_2,\eps^G) &=
\left(\eps^L_1\del\eps^L_2-\del\eps^L_1\eps^L_2\right)\del\eps^G-\frac{1}{2}\eps^G\del\left(\eps^L_1\del\eps^L_2-\del\eps^L_1\eps^L_2\right) 
\\
&\phantom{=}-\eps^L_1\del\left(\eps^L_2\del\eps^G-\frac{1}{2}\eps^G\del\eps^L_2 \right) +\frac{1}{2}\left(\eps^L_2\del\eps^G-\frac{1}{2}\eps^G\del\eps^L_2 \right)\del\eps^L_1
\\
&\phantom{=}+\eps^L_2\del\left(\eps^L_1\del\eps^G-\frac{1}{2}\eps^G\del\eps^L_1 \right) -\frac{1}{2}\left(\eps^L_1\del\eps^G-\frac{1}{2}\eps^G\del\eps^L_1 \right)\del\eps^L_2
\\
&=0.
\\ \\
\mathcal{J}_3(\eps^L, \eps^G_1,\eps^G_2) &=
\frac{1}{2}\left(\eps^L\del\eps^G_1-\frac{1}{2}\eps^G_1\del\eps^L\right)\eps^G_2- \frac{1}{2}\left(\eps^L\del\eps^G_2-\frac{1}{2}\eps^G_2\del\eps^L\right)\eps^G_1 
\\
&\phantom{=}+\frac{1}{2}\eps^G_1\eps^G_2\del\eps^L-\del\left(\frac{1}{2}\eps^G_1\eps^G_2\right)\eps^L
\\
&=0.
\\ \\
\mathcal{J}_3(\eps^G_1,\eps^G_2,\eps^G_3) &=
\frac{1}{2}\eps^G_1\eps^G_2\del\eps^G_3-\frac{1}{4}\del\left(\eps^G_1\eps^G_2\right)\eps^G_3 + \text{ cyclic}
\\
&=0.
\end{align*}

\subsection{The dual super-\Linf}
As we have seen in section \ref{basicsLinf}, there are two equivalent forms of the super-\Linf algebra. Since we have found a simple nontrivial example for such an algebra, it is an interesting exercise to work out the dual description in detail. In the previous sections we have found the algebra to be a regular \Linf algebra over a super vector space. The dual picture is simply obtained by exchanging the internal $\mathbb Z_2$ Grassmann parity for the external $\mathbb Z_2$ $s$-grading and modifying the symmetry of the products as well as the \Linf relations accordingly. In practice this means that the following, previously anticommuting products become commuting according to \eqref{supergradingsymmetry}:
\eq{
&\ell_2^{\eps^L}(\eps^G_1,\eps^G_2) && =\ell_2^{\eps^L}(\eps^G_2,\eps^G_1) &&= -2 \eps^G_1 \,\eps^G_2 \,, \\
&\ell_2^L(\eps^G,G) &&= \ell_2^L(G,\eps^G) &&= -\frac{3}{2}\del\eps^G G - \frac{1}{2}\eps^G\del G  \,,
}
with $(n,s)=(0,1)$ and $(-1,1)$ the degree of the now Grassmann even $\epsilon^G$ and $G$.
The degree of the bosonic fields $\eps^L$ and $L$ is $(0,0)$ and $(-1,0)$.
Here it also becomes clear how the two descriptions are equivalent. In the previous discussion both $\epsilon^G$ and $G$ were  Grassmann odd, and their $\ell_2$ maps were anti-commutative. By turning the Grassmann parity into an additional $\mathbb Z_2$ grading that only affects the symmetry of the \Linf maps, the right hand side becomes commuting. The \Linf maps get an additional sign factor when exchanging two previously Grassmann odd inputs, which turns the left hand side commutative as well.

Similarly, the super-\Linf relations differ from the \Linf relations if they involve two or more previously fermionic fields. There are six relations that must be checked again to verify the algebra,
$\mathcal{J}_2(\eps^G_1,\eps^G_2)$, 
$\mathcal{J}_3(\eps^G,\eps^L,G)$, and
$\mathcal{J}_3(\eps^G_1,\eps^G_2,x)$ with  $x\in\{L,G,\eps^L,\eps^G\}$. 
Note that the relations now include sign factors for the additional $\mathbb Z_2$ grading, i.e.
\eq{ 
\mathcal{J}_2=\ell_1(\ell_2(x_1,x_2))-\ell_2(\ell_1(x_1),x_2)+(-1)^{n_1n_2+s_1s_2}\ell_2(\ell_1(x_2),x_1)
}
and analogously for the signs in $\mathcal J_3$. In exchange one can freely commute all fields during calculations.

\begin{align*}
 \mathcal{J}_2(\eps^G_1,\eps^G_2)&=
 \frac{c}{2}\del^2\eps^G_1\del\eps^G_2 + \frac{c}{2} \del\eps^G_1\del^2\eps^G_2 +\frac{c}{6}\del^3\eps^G_1\eps^G_2+\frac{c}{6}\eps^G_1\del^3\eps^G_2 -\frac{c}{6}\del^3\left(\eps^G_1\eps^G_2\right) 
\\
&=0.
\\ \\
\mathcal{J}_3(\eps^G,\eps^L,G) &=
-\frac{3}{2}\del\left(\frac{1}{2}\del\eps^L\eps^G-\eps^L\del\eps^G\right)G-\frac{1}{2}\left(\frac{1}{2}\del\eps^L\eps^G-\eps^L\del\eps^G\right)\del G
\\
&\phantom{=}-\del\eps^L\left(3\del\eps^G G+\eps^G\del G\right)-\frac{1}{2}\eps^L\del\left(3\del\eps^G G+\eps^G\del G\right)
\\
&\phantom{=}+\frac{3}{2}\del\eps^G\left(\frac{3}{2}\del\eps^L G+\eps^L\del G\right)+\frac{1}{2}\eps^G\del\left(\frac{3}{2}\del\eps^L G+\eps^L\del G\right)
\\
&=0.
\\ \\
\mathcal{J}_3(\eps^G_1,\eps^G_2,L) &=
-4\del\left(\eps^G_1\eps^G_2\right)L-2\eps^G_1\eps^G_2\del L +3\del\eps^G_2\eps^G_1 L 
\\
&\phantom{=}+\eps^G_2\del\left(\eps^G_1L\right)+3\del\eps^G_1\eps_2^G L+\eps^G_1\del\left(\eps^G_2 L\right)
\\
&=0.
\\ \\
\mathcal{J}_3(\eps^G_1,\eps^G_2,G) &=
-3\del\left(\eps^G_1\eps^G_2\right)G -2 \eps^G_1\eps^G_2\del G + 3\del\eps^G_1\eps^G_2G
\\
&\phantom{=}+\eps^G_1\eps^G_2\del G + 3\eps^G_1\del\eps^G_2G+\eps^G_1\eps^G_2\del G
\\
&=0.
\\ \\
\mathcal{J}_3(\eps^G_1,\eps^G_2,\eps^L) &=
-2\left(\eps^G_1\eps^G_2\del\eps^L-\del\left(\eps^G_1\eps^G_2\right)\eps^L\right)
-2\left(\eps^L\del\eps^G_1-\frac{1}{2}\del\eps^L\eps^G_1\right)\eps^G_2
\\
&\phantom{=}-2\left(\eps^L\del\eps^G_2-\frac{1}{2}\del\eps^L\eps^G_1\right)\eps^G_1
\\
&=0.
\\ \\
\mathcal{J}_3(\eps^G_1,\eps^G_2,\eps^G_3) &=
\del\eps^G_1\eps^G_2\eps^G_3+\eps^G_1\del\eps^G_2\eps^G_3-2\eps^G_1\eps^G_2\del\eps^G_3 +\text{ cyclic}
\\
&=0.
\end{align*}

As expected, the products that were read off from the symmetry algebra indeed satisfy both definitions of super-\Linf algebras.

\section{Conclusion}

In this letter, in the context of two-dimensional conformal field
theories,  we explored the possibility of extending the \Linf
structure to super-${\cal W}$ algebras.
The question was whether such a generalization exists, and if so
whether it does require the  introduction of super-\Linf algebras.
As a simple example to clarify this issue
we employed  the super-Virasoro algebra but expect that 
a more involved  analysis of genuine nonlinear super-${\cal W}$ algebras
will confirm our findings.

We found that both the bosonic and fermionic symmetry parameters and
the bosonic and fermionic fields carry the same degree in each
case. The symmetry variations of the fields and
their closure algebra are correctly described by an ordinary
\Linf  algebra over a super vector space. We have seen how the description is equivalent to a super-\Linf algebra with only Grassmann even fields and modified symmetries and algebra relations. This is in many ways similar to the superspace approach to supersymmetric field theories and might simplify calculations in more complicated cases.

\newpage

\bibliographystyle{utphys}
\bibliography{references}
 
\end{document}